\documentclass[preprint]{article}
\pdfoutput=1

\usepackage[nonatbib]{nips_2018}
\usepackage{graphicx}
\usepackage{makecell}

\usepackage[utf8]{inputenc} 
\usepackage[T1]{fontenc}    
\usepackage{hyperref}       
\usepackage{url}            
\usepackage{booktabs}       
\usepackage{amsfonts}       
\usepackage{nicefrac}       
\usepackage{microtype}      

\title{Graph Convolutional Neural Networks for Polymers Property Prediction}
\author{
  Minggang Zeng \\
  Deep Learning Department, Institute for Infocomm Research,\\
   A*STAR (Agency for Science, Technology and Research)\\
  1 Fusionopolis Way, $\sharp$21-01 Connexis, Singapore 138632\\
\And
  Jatin Nitin Kumar \\
  Soft Materials, Institute of Materials Research and Engineering\\
  2 Fusionopolis Way, $\sharp$08-03 Innovis, Singapore 138634\\
\And
  Zeng Zeng \\
  Deep Learning Department, Institute for Infocomm Research,\\
  A*STAR (Agency for Science, Technology and Research)\\
1 Fusionopolis Way, $\sharp$21-01 Connexis, Singapore 138632\\
\And
  Ramasamy Savitha \\
  Deep Learning Department, Institute for Infocomm Research,\\
  A*STAR (Agency for Science, Technology and Research)\\
1 Fusionopolis Way, $\sharp$21-01 Connexis, Singapore 138632\\
\And
  Vijay Ramaseshan Chandrasekhar \thanks{equal corresponding author} \\
  Deep Learning Department, Institute for Infocomm Research,\\
  A*STAR (Agency for Science, Technology and Research)\\
1 Fusionopolis Way, $\sharp$21-01 Connexis, Singapore 138632\\
  \texttt{vijay@i2r.a-star.edu.sg} \\
\And
  Kedar Hippalgaonkar \textsuperscript{$\ast$} \\
  Electronic Materials, Institute of Materials Research and Engineering\\
  2 Fusionopolis Way, $\sharp$08-03 Innovis, Singapore 138634\\
  \texttt{kedarh@imre.a-star.edu.sg} \\
  }

\begin{document}

\maketitle

\begin{abstract}
A fast and accurate predictive tool for polymer properties is demanding and will pave the way to iterative inverse design. In this work, we apply graph convolutional neural networks (GCNN) to predict the dielectric constant and energy bandgap of polymers. Using density functional theory (DFT) calculated properties as the ground truth, GCNN can achieve remarkable agreement with DFT results. Moreover, we show that GCNN outperforms other machine learning algorithms. Our work proves that GCNN relies only on morphological data of polymers and removes the requirement for complicated hand-crafted descriptors, while still offering accuracy in fast predictions.

\end{abstract}

\clearpage
\section{Introduction}
Polymers are materials with tunable structures and chemical functionality that influence their physical and chemical properties.\cite{Gregory2012}  A deep understanding of the relationship between structure and properties is required for innovating novel materials. However, this relationship is complex and often not well understood.\cite{CarraherJr2012} Density Functional Theory (DFT) is useful in estimating bulk polymer properties, but is computationally expensive. It has been applied on small molecules, and a 20,000 size dataset was generated as a part of the Harvard Clean Energy Project \cite{Hachmann2011}. Using a subset from this data, by representing molecules as graphs (with individual atoms as vertices and bonds as edges), a neural fingerprinting approach expanding upon a Simplified Molecular-Input Line-Entry System (SMILES) enabled high predictive capability of solubility, drug efficacy and photovoltaic efficiency \cite{Duvenaud2015Convolutional}. Following this, a reinforced adversarial neural computer based on reinforcement learning allowed access to a broad chemical space towards predictive synthesis of small molecules \cite{Putin2018}. Moving beyond small molecules, to enable property prediction for bulk polymers, high-throughput DFT calculations were performed on different repeat units to provide a “fingerprint”, from which the final polymer properties were derived algorithmically.\cite{mannodi2016machine} Bypassing DFT for fingerprinting and instead using direct morphological information would be a significant advance in the discovery of new materials and shorten the development time for material innovation. To achieve this, a strong understanding of the relevant and most useful material descriptors influencing the polymer properties is required.

Describing the polymer in its most basic form typically constitutes identifying the atoms that make up the material as well as their arrangement with respect to each other, determined by the type of bonding that holds the material together. For inorganic crystals, where long range order is necessary, the material can be described as a lattice and a basis, where the lattice reflects the symmetry of the crystal structure, while the basis is the repeating unit. Polymers, on the other hand, are largely amorphous. This is because they do not exhibit such long range order as they do not conform to any lattice unless specifically phase controlled. However, certain polymeric properties are intrinsic to the monomer unit and translate well to actual applications. This is especially true for ground state properties such as the dielectric constant and the polymer bandgap. The dielectric constant is a reflection of the electronic polarizability of the polymer and is a consequence of the bonding nature between the constituent atoms. Typically, the bandgap depends upon the strength of the bonding. Therefore, one would expect that knowing the atoms that constitute the monomer as well as how they bond with each other - resulting in a fingerprint morphological character - is sufficient information to predict these properties. Such information is uniquely contained in a polymer’s Crystallographic Information File (CIF), which can then be converted to a two-dimensional (2D) graph and used as an input into a convolutional neural network to predict the above-mentioned properties.

In addition to morphological character intrinsic to the polymer, typically, environmental considerations are also important. Their physio-chemical environment is a complex multi-variable parameter space; intra and inter chain effects due to multi-valency and chain coiling affects functional properties, but is expected to be less influential for ground state properties. More complex descriptors that have been used in complementary approaches involve a combination of detailed atomic and morphological information along with environmental interactions.\cite{kim2018polymer}  In our work, we predict the dielectric constant and the bandgap of a large class of polymer compounds using the elegant graph convolutional neural network (GCNN) and find that the predictions are better than those where complex descriptors have been used in the recent past as illustrated in Figure 1. Specifically, by comparing to traditional descriptors, our mean absolute errors are lower than those obtained by other machine learning techniques. This clarifies that to achieve speed and accuracy in predictions, we only need to consider the morphological character of polymers.

\begin{figure}
  \centering
  \includegraphics[width=1.0\textwidth]{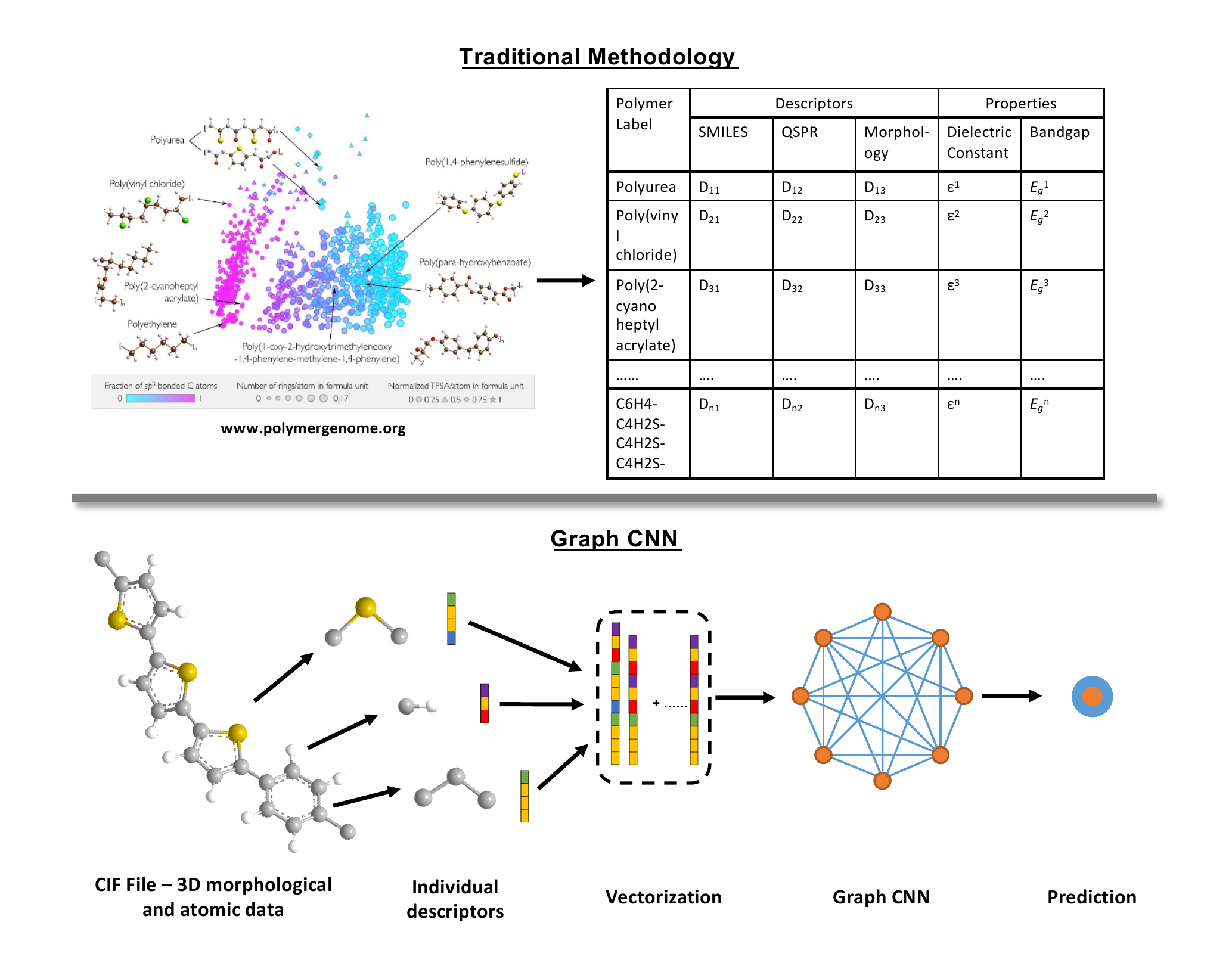}\\
  \caption{Schematic comparison between GCNN and other machine learning algorithms. The traditional methodology relies on hand engineered features crafted from SMILE string, quantitative structure-property relationship and morphology of polymers. One remarkable advantage of graph CNN is its ability to automatically learn the chemical environment of polymers and map polymer structure to abundant feature vectors for a fast and accurate prediction on polymer properties.
}
\end{figure}

\section{Dataset}

To investigate the accuracy of GCNN on predicting physical
properties of polymers, we use the publicly available dataset at the Polymergenome Project.\cite{huan2016polymer,Mannodi-Kanakkithodi2017,Mannodi-Kanakkithodi2016,mannodi2017scoping,kim2018polymer} The main purpose of this dataset is to develop polymers for energy storage and electronics applications. The dataset covers 1073 polymers that can be classified as organic or organometallic which can be further divided into three subsets according to their sources. The first subset has 34 common polymers that have been synthesized in experiments, like polyethylene, polyureas, polythioureas, polyesters, \emph{etc.}. The second subset adopted from the Crystallography Open Database (COD) contains another 253 organic and organometallic polymers. The last subset generated from computational methods includes 314 organic polymers and 472 organometallic polymers. The polymer building blocks in this subset include organic –CH2–, –NH–, –CO–, –O–, –CS–, –C$_{6}$H$_{4}$–, and –C$_{4}$H$_{2}$S–, as well as metal-containing inorganic blocks, such as –COO–Sn(CH$_{3}$)$_{2}$–OCC–, –SnF$_{2}$–, and –SnCl$_{2}$–. The polymer building block repeats along the 3-dimensional crystal axis, and the repeat unit with symmetry information is fed into the GCNN.

\section{Experimental Setup}
\subsection{Input}
To build a regression model, the input of GCNN includes CIF files recording the structure of the polymers, the target properties for each polymer and a JSON file that stores the initialization vector for each atom.
\subsection{Model Architecture}
A polymer graph can be represented by nodes and edges using the atomic feature vector $\mathbf{v}_{i}$ and the bonding feature vector $\mathbf{u}_{(i;j)_{k}}$, respectively.\cite{Duvenaud2015Convolutional,Gilmer2017Neural,xie2018hierarchical,Xie2018Crystal} These vectors are obtained by one hot encoding. With the help of a non-linear convolution function, the multiple convolutional layers automatically learn the atomic features after iterating through surrounding chemical features (atoms/bonds) and different convolutional layers. A pooling layer of normalized summation is then used to generate an overall feature vector $\mathbf{v}_{c}$.
To improve the performance of the GCNN, two fully connected hidden layers
are added to capture the complicated structure-property relationship. Finally, an output layer connected to the top hidden layer is used to predict the physical property of polymers.

\subsection{Hyperparameter Optimization}
The database is randomly divided into training, validation
and test sets with the ratio of 6:2:2. The network parameters
are optimized via Stochastic Gradient Descent (SGD), and the optimum hyperparameters are determined by the lowest mean absolute error (MAE) in the validation set using the DFT result as the ground truth.

\begin{table}
  \caption{Hyperparmeters for different algorithms}
  \label{Tab1}
  \centering
  \begin{tabular}{ll}
    \toprule
    Method & Hyperparameters\\
    \midrule
    GCNN for $\varepsilon$ (E$_{g}$) & \makecell[l]{Learning rate: 0.001 (0.01),
    momentum: 0.9 (0.8), \\
    hidden-feature-length: 256 (64),
    weight decay: 1e-8 (1e-5),\\
    atomic-feature-length: 32 (32),\\
    number of CNN layers: 2 (5), \\
    number of hidden layer: 2 (5)} \\
    Kernel Regression & \makecell[l]{alpha: 1e-05,
     gamma: 1.25e-09,
     kernel: rbf} \\
    Random Forest & \makecell[l]{min\_samples\_leaf: 10, n\_estimators: 150,
    oob\_score: True} \\
    Gradient Boosting & \makecell[l]{alpha: 0.7,
    learning\_rate: 0.1,
    max\_depth: 5} \\
    Neural Network & \makecell[l]{alpha: 0.001,
    hidden\_layer\_sizes: 100,
    momentum: 0.7} \\
    \bottomrule
  \end{tabular}
\end{table}

\section{Predictive Performance}
We ran two experiments to demonstrate that GCNN can obtain comparable accuracy to DFT for polymers. Here we focus on the bandgap ($E_{g}$) and  the dielectric constant ($\varepsilon$). These two properties are important in order to screen polymer materials regardless of specific applications. The DFT results of $E_{g}$ and $\varepsilon$ are used as the ground truth. They are obtained with hybrid electron exchange-correlation functionals and density functional perturbation theory, respectively.\cite{huan2016polymer} The optimized hyperparameters of GCNN for $E_{g}$ and $\varepsilon$ are listed in Table~1. Figures 2(a,b) compare the predictive performance of GCNN versus the DFT ground truth. Impressive agreement with DFT is found in predicting the dielectric constant. A MAE value of 0.24 is achieved on the test set, which is lower than the published work using a similar dataset and Gaussian process regression.\cite{kim2018polymer} Given the error of DFT calculations compared with experiments and the small MAE obtained in our study, GCNN may achieve accuracy in predicting polymer properties as compared to experiments. Comparatively, a higher MAE (0.41 eV) is found using GCNN to predict the energy bandgap of the polymer dataset. We illustrate the data distribution for $E_{g}$ and $\varepsilon$ in Figure 2(c). The dielectric constant of the dataset is mainly concentrated in the region of $\varepsilon = 3$, with the mean and variation of 3.47 and 0.92, respectively. Comparatively, the $E_{g}$ is much more dispersed (mean = 4.417 eV, variation = 2.75 eV) with an obvious tail weighted in the high bandgap range. Therefore, the MAE is high since lesser data is available for the polymer in this energy bandgap tail region. The inset of Figure~2(b) shows the MAE for $E_{g}$ as the function of dataset size. We find a systematic decrease in MAE with increasing the size of dataset from 128, 256, 512 to 1024. This indicates the prediction of $E_{g}$ by GCNN can be improved if more polymer data are provided.\cite{kim2018polymer}.

\begin{figure}
  \centering
  \includegraphics[width=1.0\textwidth]{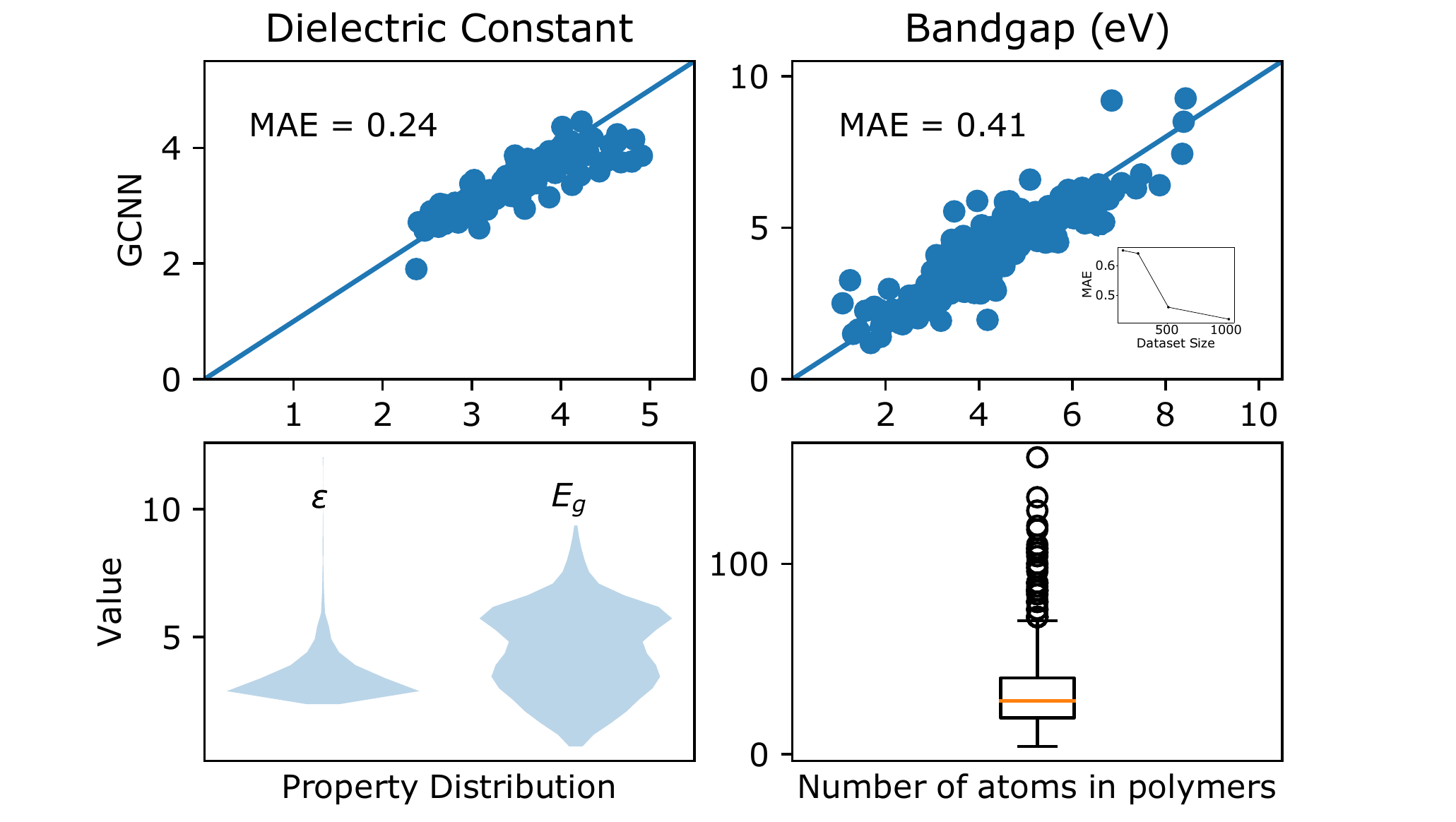}\\
  \caption{(a,b) The predicted dielectric constant and bandgap by GCNN compared with the DFT results, showing that the GCNN is an accurate predictive tool. The inset of (b) shows a systematic decrease in MAE with increasing the size of dataset from 128, 256, 512 to 1024. (c) Value distribution showing a narrow range for the dielectric constant and a wider range for the bandgap. (d)Statistics showing the large range in the number of atoms in each polymer; the GCNN does not discriminate against large polymers and is still able to achieve good prediction.}
\end{figure}

\section{Comparison with other Machine Learning methods}
Besides GCNN, where we only use the cif and the atomic .json files as inputs, we also use other Machine Learning (ML) methods to establish a regression model for the polymer structure-property relationship. The feature extraction of the polymer dataset is implemented with the Matminer package.\cite{ward2018matminer} Table~3 lists the feature extraction modules that we apply, which produces 158 individual feature descriptors. These hand-crafted features reflect some key physical and chemical properties of polymer systems, such as structural heterogeneity and chemical ordering.
Based on these physically relevant descriptors, we apply a series of machine learning algorithms using the Scikit-learn package to study the structure-property relationship of the polymer dataset.
The ML models have internal 4-fold cross-validation to minimise over-fitting and ensure model generality.
The models are trained using SGD with the ADAM optimizer; and the GridSearchCV method in Scikit-Learn is applied to tune hyperparameters. The optimized hyperparameters are listed in Table~1; and the MAE values of these ML algorithms are listed in Table~2. It can be seen that GCNN outperforms all these ML algorithms in predicting the $E_{g}$ and $\varepsilon$ within this polymer dataset. This indicates that the feature vectors generated from the GCNN function better in predicting the physical properties of polymers. This may come from the fact that GCNN feature vectors take into account global spatial geometry, as well as accurate local atomic configuration, as listed in Table~3. As shown by the statistic analysis of the polymer dataset in Figure~2(d), GCNN works well even on long polymer chains, regardless of a simplified graphical representation. Our results suggest that GCNN has the potential to serve as an excellent forward predictive model for polymers.

\begin{table}
  \caption{Summary of the prediction performance of three different properties by different ML methods}
  \label{Tab1}
  \centering
  \begin{tabular}{cccccc}
    \toprule
    \multicolumn{4}{r}{MAE}\\
    \cmidrule{2-6}
    Property & GCNN & Kernel Regression & Random Forest & Gradient Boosting & Neural Network\\
    \midrule
     $\varepsilon$ & 0.24 & 0.425 & 0.355 & 0.359 & 0.59\\
     $E_{g} (eV)$ & 0.41 & 0.652 & 0.505 & 0.446 & 0.509\\
    \bottomrule
  \end{tabular}
\end{table}

\begin{table}
  \caption{Comparison between the GCNN generated feature classes and the hand-crafted feature classes obtained via the Matminer package. The Kernel Regression, Random Forest, Gradient Boosting and Neural Network algorithms use these listed hand-crafted feature classes.}
  \label{Tab3}
  \centering
     \begin{tabular}{lll}
      \toprule
    Hand-crafted feature classes  & GCNN generated feature classes \\
     \midrule
    Structural Heterogeneity & Group number \\
    Chemical Ordering & Period number \\
    Maximum Packing Efficiency & Electronegativity \\
    Stoichiometry & Covalent radius \\
    Element Property & Valence electrons \\
    Valence Orbital & First ionization energy \\
    Ionic Property & Electron affinity \\
      & (s,p,d,f) Block \\
      & Atomic volume \\
      & Atomic distance\\
    \bottomrule
    \end{tabular}%
\end{table}

\section{Discussion and Future Work}
Given the strength of the GCNN predictions, it is good to note that the present approach has two limitations. The first is that the ground truth is based on simulations rather than experimental data, which is a result of the difficulty of performing high-throughput experiments and hence sparse experimental data. The second is that the present technique only allows for the prediction of bulk polymeric properties as opposed to polymer-solvent interactions and composites. While the first limitation is due to lack of data availability, the second could be addressed by investigating feature importance using the GCNN to allow for interpretability. We seek inspiration from a recent work on developing a methodology of vectorizing individual molecular descriptors via multi-dimensional correlation to account for such complexities, thereby enabling the discovery of novel functional molecules.\cite{Gomez-Bombarelli2018Automatic} This will then allow us to build an accurate yet fundamental design-property understanding which in turn could predict experimental polymer behaviour. Moreover, this sort of understanding could also extend the capability beyond bulk polymers to polymer composites and polymers in solution. Most importantly this strong understanding of polymers will allow us to predict the required chemical and structural design of a polymer based on its physical property requirement - the problem of inverse design – which would lead to huge impact in both industry and academia.

\section{Conclusion}
We applied several machine learning algorithms to predict dielectric constant and bandgap from a large dataset of crystallography data of polymers. Our models included graph convolution neural network, random forest, kernel regression, gradient boosting and conventional neural network, where the lowest mean absolute error for both properties were reported for GCNN. These results indicate that GCNN offers an effective approach for fast and accurate prediction of polymer properties starting from the atomic and morphological character of polymer. This conceptual advance allows us to move into the realm of faster predictions, relying on a smaller amount of metadata, and paves the way for inverse design, where polymers can be designed with a final property in mind.

\bibliographystyle{unsrt}
\bibliography{NIPSW2018_GCN_Polymer_V3}
\end{document}